\documentclass[12pt,preprint]{aastex}
\shorttitle{Mass Loss Histories of AFGL 2688, OH 231.8+4.2, \& IRAS 16342$-$3814}

\begin{document}

\title{A Spitzer Study of the Mass Loss Histories of Three Bipolar Pre-Planetary Nebulae}

\author{Tuan Do, Mark Morris}
\affil{Physics and Astronomy Department, University of California,
	Los Angeles, CA 90095-1547}

\author{Raghvendra Sahai and Karl Stapelfeldt}
\affil{Jet Propulsion Laboratory, California Institute of Technology,
	Pasadena, CA 91109}

\begin{abstract}
We present the results of far-infrared imaging of extended regions around three bipolar pre-planetary nebulae, AFGL 2688, OH 231.8+4.2, and IRAS 16342$-$3814, at 70 and 160 $\mu$m with the MIPS instrument on the Spitzer Space Telescope.  After a careful subtraction of the point spread function of the central star from these images, we place constraints on the existence of extended shells and thus on the mass outflow rates as a function of radial distance from these stars.  We find no apparent extended emission in AFGL 2688 and OH 231.8+4.2 beyond 100 arcseconds from the central source.  In the case of AFGL 2688, this result is inconsistent with a previous report of two extended dust shells made on the basis of ISO observations.  We derive an upper limit of $2.1\times10^{-7}$ M$_\odot$ yr$^{-1}$ and $1.0\times10^{-7}$ M$_\odot$ yr$^{-1}$ for the dust mass loss rate of AFGL 2688 and OH 231.8, respectively, at 200 arcseconds from each source. In contrast to these two sources, IRAS 16342$-$3814 does show extended emission at both wavelengths, which can be interpreted as a very large dust shell with a radius of $\sim$ 400 arcseconds and a thickness of $\sim$ 100 arcseconds, corresponding to 4 pc and 1 pc, respectively, at a distance of 2 kpc.  However, this enhanced emission may also be galactic cirrus; better azimuthal coverage is necessary for confirmation of a shell.  If the extended emission is a shell, it can be modeled, with some assumptions about its dust properties, as enhanced mass outflow at a dust mass outflow rate of $1.5\times10^{-6}$ M$_\odot$ yr$^{-1}$ superimposed on a steady outflow with a dust mass outflow rate of $1.5\times10^{-7}$ M$_\odot$ yr$^{-1}$. Because of the size of the possible shell, it is likely that this shell has swept up a substantial mass of interstellar gas during its expansion, so these estimates are upper limits to the stellar mass loss rate.  We find a constant color temperature of 32 K throughout the circumstellar envelope of IRAS 16342$-$3814, which is consistent with heating by the interstellar radiation field.
\end{abstract}


\section{Introduction}

Pre-planetary nebulae (PPNe) represent a fleeting stage of stellar evolution. These transitional objects arise between the rapid mass loss phase at the end of the asymptotic giant branch (AGB) and the ionized planetary nebula stage. AGB stars are important because they are the Galaxy's main mechanism for the replenishment of dust and gas into the interstellar medium (ISM), ejecting most of their main sequence mass within a few times $10^5$ years because of their high mass loss rates. By looking at the circumstellar material around PPNe, we can see an imprint of the activity of the earlier AGB. PPNe are also in a stage where the geometry of the mass loss from the central star usually changes from spherically symmetric to an axially symmetric bipolar outflow, as illustrated in the HST images of AFGL 2688 \citep{1998ApJ...492L.163S}. In these images, a spectacular bipolar structure is superimposed on a set of concentric shells of gas and dust ejected near the end of the AGB. In the study reported here, we have attempted to characterize the mass loss behavior of the progenitor AGB stages of three systems that are now bipolar. 

Using the direct detection of infrared emission from dust lost over as much as $10^5$ years during the AGB, it is possible to study the mass loss histories of these objects. There has been a dearth of observations of large extended shells of evolved stars out to the distances sampled in this project primarily because of the difficulty in observing in the far infrared where the cool dust has its peak emission. With the Spitzer Space Telescope \citep{2004ApJS..154....1W}, the far infrared is now accessible at a resolution and sensitivity sufficient to potentially resolve the structures of these dust shells, and thereby to test theories of mass loss in AGB stars. 

Several tantalizing cases were observed by IRAS, showing that large dust shells exist around some evolved stars. For example, \citet{1986ApJ...310..842G} found that R CrB shows a very large shell with a radius of 4 parsecs, and \citet{1990A&A...229L...5H} reported a dust shell with a diameter of 30 - 40 arcminutes ($\sim 1$ pc) around W Hya. However, these observations could not resolve the structure that may be present in the shells, or address the possibility of multiple shells. If the mass loss during the AGB is constant, we should expect to see a smooth envelope with column density declining as $1/b$, where $b$ is the displacement from the star in the sky. However, if the mass loss rate fluctuates as the star goes through thermal pulses during the AGB, as the models suggest \citep[e.g.,][]{1993ApJ...413..641V}, then we should see enhanced emission in the form of multiple dust shells around our objects.  For example, \citet{2000ApJ...545L.145S} reported on the basis of ISO data that there are large concentric shells in AFGL 2688 and AFGL 618. This result is among a small set of observations showing evidence for periodic mass loss caused by thermal pulsation as well as the time scale between thermal pulses predicted by AGB evolutionary models \citep[e.g.,][]{1993ApJ...413..641V}. Confirmation of these shells at higher resolution and greater sensitivities motivated our selection of AFGL 2688 as a target for this study. There have been no previous claims of large-scale dust emission from OH 231.8+4.2 (hereafter OH 231.8) or IRAS 16342$-$3814. However, they are presently very young PPNe with high mass loss rates of approximately $10^{-4} M_{\odot} ~ yr^{-1}$ \citep{2001A&A...373..932A, 1999ApJ...514L.115S}, which suggests that the emission from dust produced during the AGB may be readily visible. 

Here we report on the lack of very extended emission, spherically symmetric or otherwise, from AFGL 2688 and OH 231.8. However, we do see possible evidence for a very large diffuse dust shell around IRAS 16342$-$3814 with a radius of 400 arcseconds, though this extended emission might also be from galactic cirrus. 

\section{Observations and Data Reduction}

AFGL 2688 (the Egg Nebula), OH 231.8, and IRAS 16342$-$3814 were observed with the MIPS instrument \citep{2004ApJS..154...25R} at 70 and 160 $\mu$m on the Spitzer Space Telescope. These PPNe were observed along two mutually perpendicular scan paths in order to sample dust emission out to about 800 arcseconds from the central source, and to determine background levels. At 70 $\mu$m, the scan paths are $15^\prime \times 3.0^\prime$ in one direction and $11^\prime \times 7.7^\prime$ in the other, while at 160 $\mu$m, the scan paths are $15^\prime \times 2.6^\prime$ and $10.5^\prime \times 6.5^\prime$. The pixel scale is 9.2 arcseconds at 70 $\mu$m and 16 arcseconds at 160 $\mu$m. However, the central sources were not directly observed in order to avoid persistence artifacts from these bright sources. This observing strategy limits the azimuthal coverage of the sources, but should give an adequate estimate of the presense of well-defined shells or asymmetries such as large-scale bipolarity. 

\begin{figure}
	\centering
	\includegraphics[scale=.75]{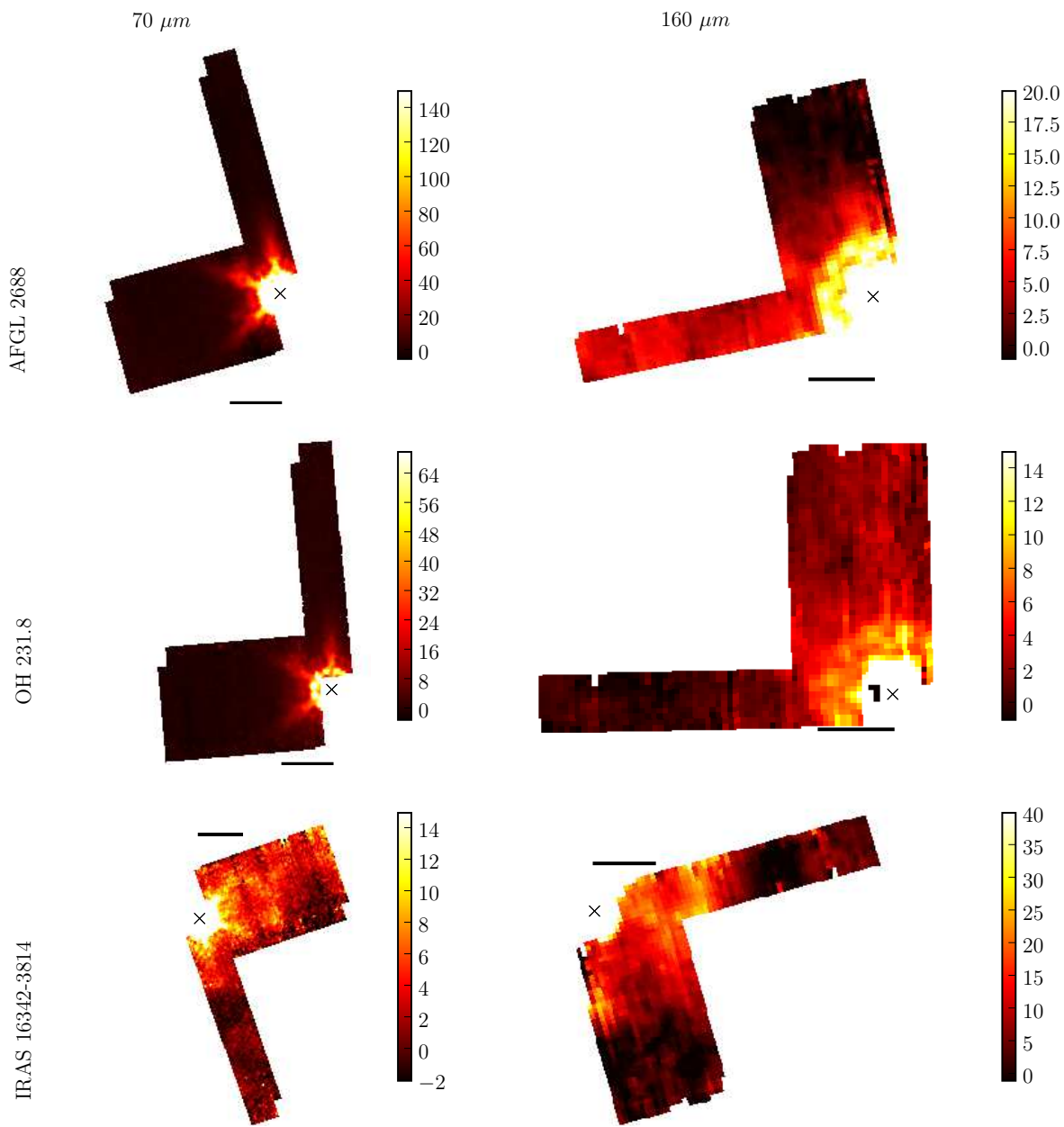}
	\caption{The three PPNe in our sample after mosaicking and background subtraction.  The units of the colorbars are in MJy Sr$^{-1}$. The bars in the images are 200$^{\prime\prime}$  in length for scale and the $\times$ marks the location of the point source. Top: AFGL 2688. Center: OH 231.8. Bottom: IRAS 16342$-$3814. Note the prominence of the PSF features at 70 $\mu$m in all the images. At 160 $\mu$m, the PSF is not apparent in IRAS 16342$-$3814. North is up and east is to the left.}
	\label{fig:initial_mosaic}
\end{figure}

The basic science calibrated data (BCD) was reduced using the software package MOPEX\footnote{http://ssc.spitzer.caltech.edu/postbcd/} from the Spitzer Science Center (SSC). MOPEX was used for rejecting outliers as well as for co-adding and mosaicking the individual BCD frames to create the final image. Further reduction was done manually to remove striping due to bright latents as described in the MIPS data handbook\footnote{http://ssc.spitzer.caltech.edu/mips/dh/}. The correction involved finding the median value measured for every pixel from a series of BCD frames far from the central source, where we assume there to be no extended emission, then subtracting this median from each data frame before co-adding and mosaicking. The median subtraction removes both the bright latents and the uniform component of background emission. Figure \ref{fig:initial_mosaic} shows the final mosaicked images.

The most prominent features of the 70 $\mu$m images of the bipolar targets are the diffraction spikes from the point spread function (PSF) of the central source. Although we did not image the central star, the wings of the PSF are  still present and bright out to about 200 arcseconds from the source. At 160 $\mu$m, the PSF is less pronounced but we see much more background than at 70 $\mu$m from galactic cirrus. The galactic cirrus emission presents a problem in determining the true sky background since it covers much of our fields. For both the 70 and 160 $\mu$m images, we use the median surface brightness of regions of low and uniform brightness far from the central source as an estimate of the sky background. This procedure removes a uniform background from the images, retaining possible enhanced emission from the source and galactic cirrus. The uncertainty in the background (see below) at 160 $\mu$m is higher than at 70 $\mu$m because it is more difficult to find a large patch of uniform surface brightness to estimate the sky background.

\subsection{Sensitivities}

The sensitivity of our data is estimated by examining the standard deviation of a patch of uniform background far from the source both before and after the median background subtraction method described above. For AFGL 2688, before background subtraction, the mean of a patch in the eastern scan path at 70 $\mu$m is 14.1 MJy Sr$^{-1}$ with a standard deviation of 1.11 MJy Sr$^{-1}$. Some of the variance in the background is likely from the effects of bright latents from the detector pixels, which are removed by the median background subtraction described above. After median subtraction, the mean background is at 0.6 MJy Sr$^{-1}$ with a standard deviation of 0.92 MJy Sr$^{-1}$. The mean surface brightness at 160 $\mu$m before background subtraction is 31.45 MJy Sr$^{-1}$ with a standard deviation of 1.67 MJy Sr$^{-1}$. After background subtraction, the mean is at 0.33 MJy Sr$^{-1}$ with a standard deviation of 1.51 MJy Sr$^{-1}$. The sensitivity for OH 231.8 is better than for AFGL 2688, probably because of the lower sky background levels. The sensitivity for IRAS 16342$-$3814 is comparable to that of AFGL 2688 at 70 $\mu$m, but worse at 160 $\mu$m due to a higher sky background and greater presence of galactic cirrus. The mean background and sensitivities for all three sources at 70 and 160 $\mu$m are given in Tables \ref{tab:70_sensitivity} and \ref{tab:160_sensitivity}, respectively.

\subsection{PSF Subtraction}

\begin{figure}
	\centering
	\plotone{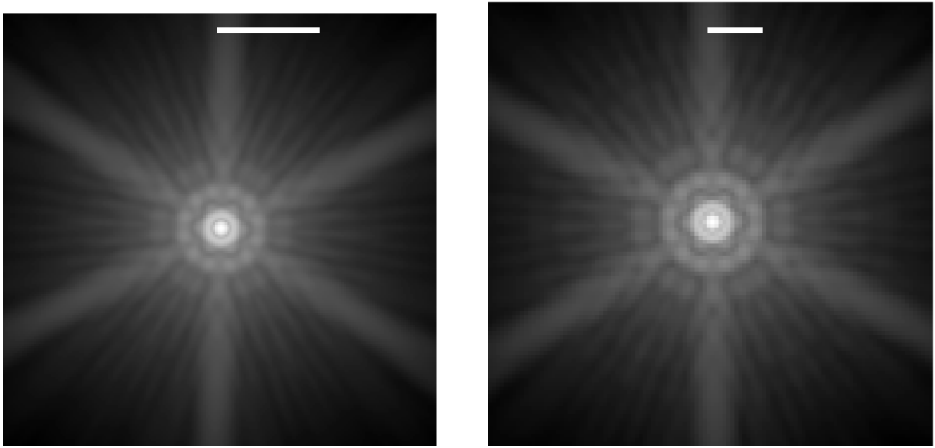}
	\caption{Model \textit{stinytim} PSFs in a log stretch to emphasize the Airy rings and diffraction spikes. Left: The 70 $\mu$m PSF. Right: The 160 $\mu$m PSF. The lines in the figures are 200 arcseconds in length. For scale, the surface brightness of the diffraction spikes $\sim$ 200 arcseconds from the center is $\sim 10^{-4}$ times lower than at the center. See Figures \ref{fig:afgl2688_psfsub} and \ref{fig:afgl2688_70_1d} for azimuthally averaged plots of the model PSF.}
	\label{fig:model_psf}
\end{figure}

Since the PSF pattern from the central source is so prominent at 70 $\mu$m, we attempted to fit a model PSF to our images in order to subtract the bright central source in search of extended dust emission near the source. The \textit{stinytim} software provided by the SSC was used to generate model PSFs.  We chose to use the default MIPS throughput curve along with a blackbody spectrum at several temperatures ranging from 20 to 100 K. See Figure \ref{fig:model_psf} for examples of the model PSFs. Multidimensional fitting was carried out for our images in order to minimize the residuals of the PSF subtraction using various x-y shifts as well as the scaling of the PSF flux values. The best fit was found by minimizing the square root of the sum of the squares of the differences in flux. We found several limitations in the model PSFs when attempting to subtract the PSF from the images. Even for the best-fit PSFs, the subtraction still leaves some obvious PSF residuals at a level of $10^{-4}$ relative to the peak as extrapolated from the model PSF. These residuals show that the wings of the model PSF are not well characterized at that flux level. The PSF subtraction also tends to leave portions of the region within the Airy ring with negative flux values. The most likely reason is a non-linear pixel response as the pixels approach saturation so that although the model has a higher brightness value closer to the core of the PSF, the actual pixel values are leveling off for our bright sources. Unfortunately, there are presently no empirical data on the MIPS 70 or 160 $\mu$m PSF to compare to the model PSF at distances greater than 100 arseconds from the peak. Beyond about 240 arseconds, PSF features are no longer present in the MIPS images so our ability to detect extended emission there is only limited by background and integration time. 

\begin{figure}
	\centering
	\plotone{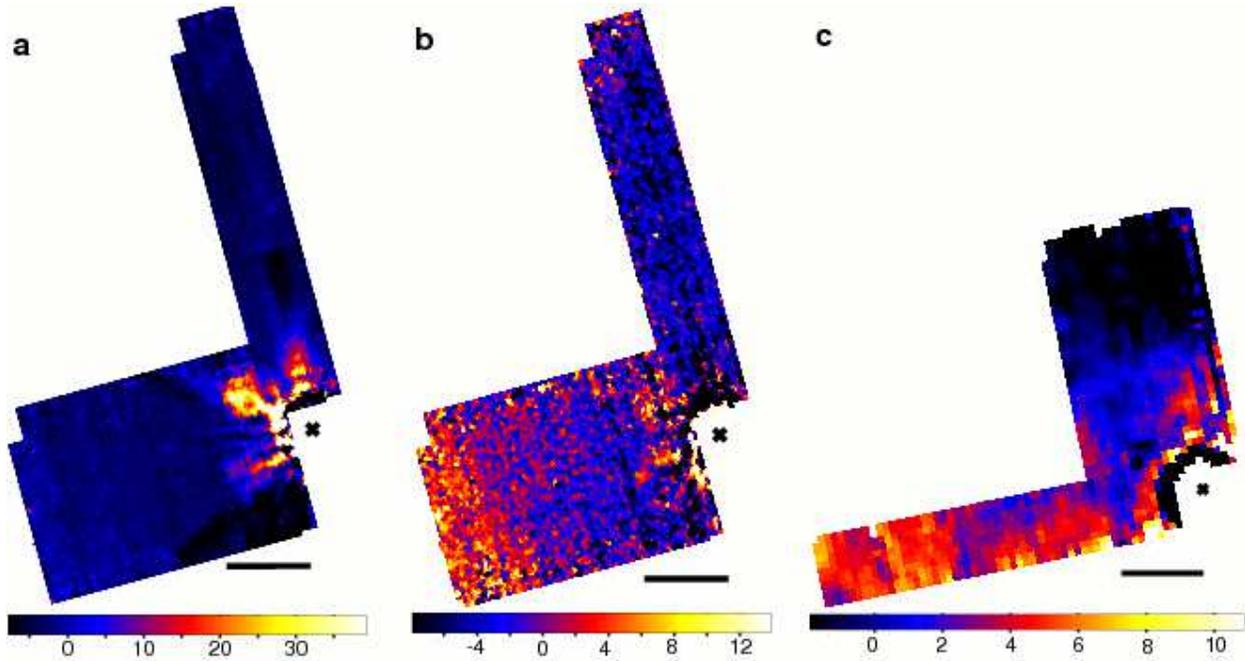}
	\caption{PSF subtracted images of AFGL 2688. The color stretches are linear, with the units in MJy Sr$^{-1}$. The x marks the location of the point source while the line represents 200$^{\prime\prime}$ for scale. a) PSF subtraction using the \textit{stinytim} model PSF at 70 $\mu$m. The three PSF diffraction spikes visible in Figure \ref{fig:initial_mosaic} appear to subtract differently, leaving different levels of residuals. b) PSF subtraction at $70 \mu m$ using OH 231.8 as a PSF, showing almost no residuals except for the region at about 60 degrees east of north. c) PSF subtraction using the \textit{stinytim} PSF at 160 $\mu$m. The feature in the northern scan path coincides with the position of the diffraction spike in the model PSF. The images have the same orientation as Figure \ref{fig:initial_mosaic}, with north being up.}
	\label{fig:afgl2688_psfsub}
\end{figure}

\begin{figure}
	\centering
	\plotone{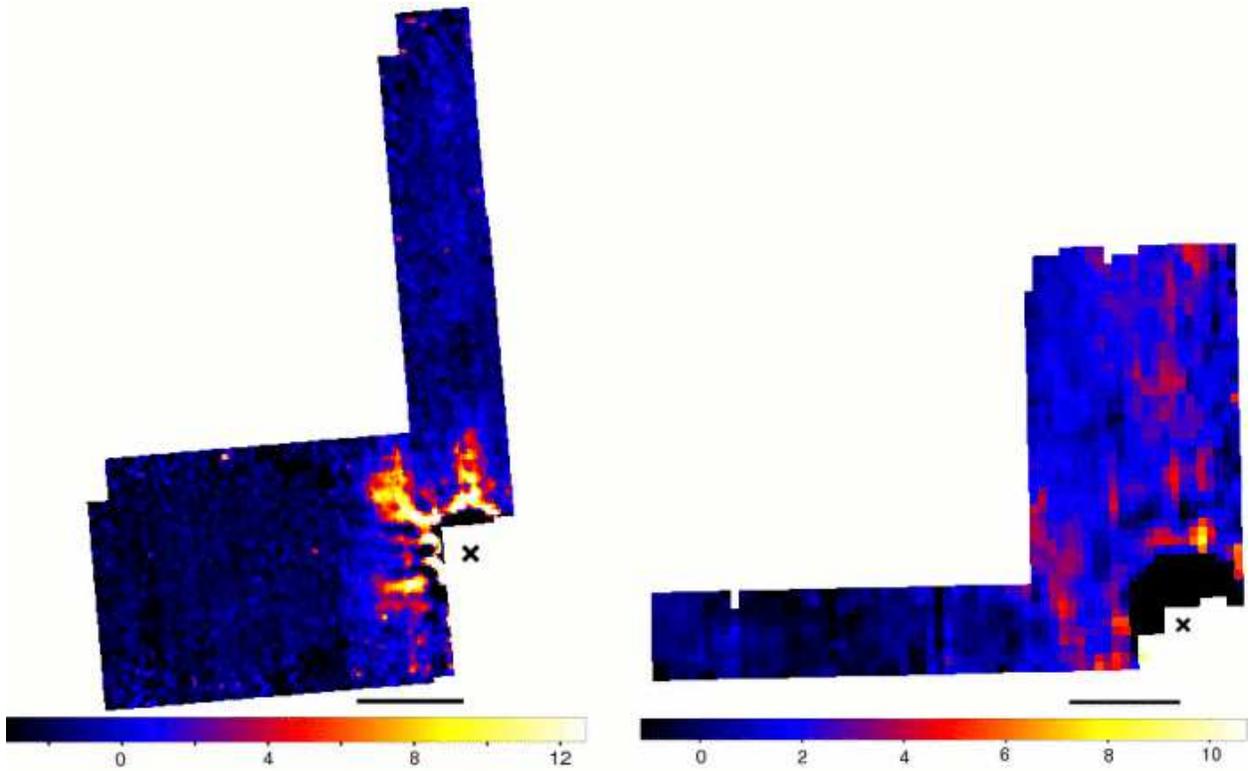}
	\caption{\textbf{Left:}  PSF subtraction of OH 231.8 using the model PSF at 70 $\mu$m. The residuals are very similar to those resulting from the PSF subtraction of AFGL 2688 in Figure \ref{fig:afgl2688_psfsub}. \textbf{Right:} PSF subtraction at 160 $\mu$m. The northern scan path has more structure than the eastern scan, probably from galactic cirrus. The units of the colorbars are in MJy Sr$^{-1}$. The x marks the location of QX Pup, the central star of OH 231.8. North is up as in Figure \ref{fig:initial_mosaic}.}
	\label{fig:oh231_psfsub}
\end{figure}

The results of PSF subtraction for AFGL 2688 are shown in Figure \ref{fig:afgl2688_psfsub}. Note the remaining features near the north and at a position angle of $\sim 50^{\circ}$ east of north, about 150 arcseconds from location of the source in the PSF subtraction in Figure \ref{fig:afgl2688_psfsub}a. This residual can also been seen in the roughly the same region of the 70 $\mu$m PSF subtraction of OH 231.8 in Figure \ref{fig:oh231_psfsub}. It seems unlikely that these features are physical, but the northern feature in AFGL 2688 does align with its bipolar outflows seen in the optical and infrared, but which are observed within the inner 10 arcseconds of the PPN. The blob in AFGL 2688 at a position angle of $\sim 50^{\circ}$ also roughly aligns with the outflow direction seen in previous CO observations \citep{1997A&A...328..290S}, also at around 10 arcseconds from the central star. The fact that the features we see are at nearly 200 arcseconds from the source and lie almost on top of the diffraction rays of the original PSF casts doubt upon their reality. This dilemma cannot be resolved until a well sampled empirical PSF out to several arcminutes is available. We may be seeing possible features in the wings of the actual PSF that the model has failed to reproduce.

As a substitute for an empirical PSF, we used OH 231.8 to subtract the PSF from AFGL 2688. Figure \ref{fig:afgl2688_psfsub}b shows the result. The residuals from using OH 231.8 as an empirical PSF appear to be much smaller than those obtained using the model PSF, reducing the residual of 30 MJy Sr$^{-1}$ to less than 10 MJy Sr$^{-1}$ at 100 arcseconds. Note that there are almost no PSF features, such as diffraction spikes, remaining in Figure \ref{fig:afgl2688_psfsub}b compared to Figure \ref{fig:afgl2688_psfsub}a. However, it may be problematic to use OH 231.8 +4.2 as an empirical PSF because it may have extended emission as well, although it would be remarkably fortuitous if both sources had extended emission having precisely the same morphology. It is also problematic to scale OH 231.8 to the same surface brightness as AFGL 2688 because in order to subtract the PSF, the whole image needs to be scaled, thus scaling the background as well. Because AFGL 2688 is over twice as bright as OH 231.8, but the background in AFGL 2688 is not, scaling to the level of the AFGL 2688 PSF would overemphasize the background in OH 231.8 so that the residuals after the PSF subtraction might be dominated by the background in OH 231.8. We can nevertheless see that the PSF subtraction using OH 231.8 results in residuals at a level $\sim$ 10 times below that from using the model PSF at 150 arcseconds, which suggests that OH 231.8 is either a point source at 70 $\mu$m or that its extended structure has the same orientation, scale, and shape as in AFGL 2688.

The two-dimensional PSF subtraction at 160 $\mu$m was done in a similar way using the model \textit{stinytim} PSF. There are PSF features remaining after this subtraction as well (see Figure \ref{fig:afgl2688_psfsub}); the ends of the diffraction spikes at 200 arcseconds from the source in the original image are not completely removed. We did not attempt a PSF subtraction of the 160 $\mu$m image of IRAS 16342$-$3814, because that the PSF of the central source is weak enough at this wavelength that its features are below the background and the extended structures. 

\subsection{1-D PSF Subtraction}

In order to investigate the hypothesis of spherical shells around a central source, we have determined an azimuthal average of the intensity around the central source. Such an average would show enhancements for spherically concentric shells projected as circularly symmetric features from episodic mass loss during the AGB phase. An azimuthal average also has the advantage that it greatly reduces the parameter space necessary to fit the model PSF. The source position and intensity are found by fitting to the brightest Airy ring visible in our images, which is the second brightest Airy ring of the PSF (see Figure \ref{fig:model_psf}), since the source itself does not appear in our images. The PSF Airy ring used for scaling the PSF is located 76 arcseconds from the source at 70 $\mu$m and 170 arcseconds from the source at 160 $\mu$m.  We also find that the azimuthal averages provide a subtraction with less residuals than using the two dimensional subtraction because localized irregularities of the PSF are averaged away. 

\begin{figure}
	\centering
        \plotone{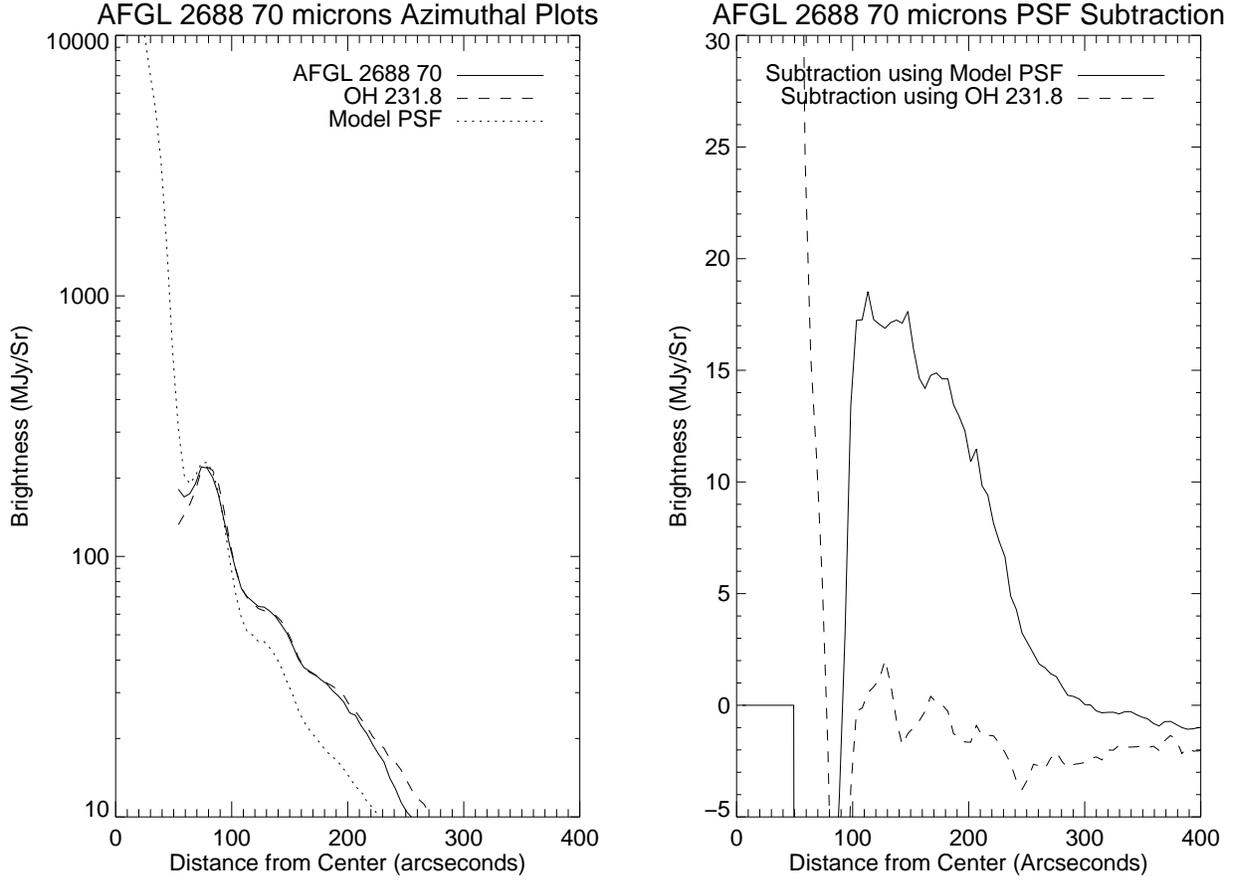}
	\caption{\textbf{Left}: The azimuthally averaged surface brightness of AFGL 2688, OH 231.8, and the model PSF at 70 $\mu$m. The model PSF and OH 231.8 are scaled to the Airy ring from the AFGL 2688 PSF at 76 arcseconds. \textbf{Right}: The result of the PSF subtractions. The surface brightness drops to background levels beyond about 250 arcseconds. The excess emission left from the PSF subtraction using the \textit{stinytim} model PSF is very similar to the residuals left after the same model PSF subtraction of OH 231.8 in Figure \ref{fig:oh231_70_1d}. The PSF subtraction using OH 231.8 as a template PSF instead of the model PSF is shown with a dashed line; the excess emission from the subtraction using the model PSF is almost completely eliminated.}
	\label{fig:afgl2688_70_1d}
\end{figure}

The one dimensional PSF subtractions for AFGL 2688 and IRAS 16342$-$3814 at 70 $\mu$m were done using both the model PSF and OH 231.8, and are shown in Figure \ref{fig:afgl2688_70_1d}. The profiles of AFGL 2688 and OH 231.8 follow each other almost exactly out to beyond 200 arcseconds from the source. This further indicates that OH 231.8 and AFGL 2688 are both strongly centrally concentrated at 70 $\mu$m, unless they both have exactly the same azimuthal excesses. They both also match the model PSF very well out to 100 arcseconds beyond which both appear to have slight excess ($<$ 1\% of the extrapolated peak of the PSF), above the model PSF. Once again, this excess in the wings is more likely due to the model PSF not reproducing the wings of the actual PSFs rather than resulting from physical emission associated with the sources. At 160 $\mu$m, we only use the model PSF for subtraction because of possible contamination due to galactic cirrus in OH 231.8.

\section{Results and Analysis}

\subsection{AFGL 2688}

\subsubsection{70 $\mu$m}

At 70 $\mu$m, the PSF-subtracted images of both AFGL 2688 and OH 231.8 are similar. The PSF residuals extend to about 150 arcseconds from the source, which may be obscuring some physical extended features. However, there are no obvious circularly symmetric residuals in the PSF subtraction even with the confusion near the central source, which would be expected if there had been spherically symmetric mass loss above a dust mass loss rate of $\sim 2.1\times10^{-7} M_\odot ~ $yr$^{-1}$ (see discussion section). Previous studies with ISO by \citet{2000ApJ...545L.145S} led to the claim of large shells around AFGL 2688 having radii of 150 and 300 arcseconds on the basis of a one-dimensional scan of the Egg Nebula with ISOPHOT at 120 and 180 $\mu$m. We find, with better spatial resolution and azimuthal coverage using Spizer, that at 70 $\mu$m, there are no signs of shell-like extended emission in our field. At 150 arcseconds from the source, the PSF subtraction residuals have a surface brightness of about 16 MJy Sr$^{-1}$ using the model PSF and about 1 MJy Sr$^{-1}$ using OH 231.8 as the empirical point source(Figure \ref{fig:afgl2688_70_1d}). Note also that there is no discernible excess emission above the fluctuations in the background ($\sigma \sim 0.92$ MJy Sr$^{-1}$) at 300 arcseconds from the source.

\subsubsection{160 $\mu$m}

The 160 $\mu$m image (see Figure \ref{fig:afgl2688_psfsub}) shows roughly uniform enhanced emission to the edge of our field of view at 1000 arcseconds from the source in the eastern scan path; the northern scan show enhanced emission out to about 300 arcseconds, then dropping off to background levels. The enhanced emission in the two paths is asymmetric as it does not fall off with distance in the eastern scan path. There does not appear to be any structure to the emission in either scan, other than that which may be attributable to the PSF diffraction spike in the north. Because the residual extended emission shows no symmetry about the central star, and because it shows no intensity falloff in the eastern scan path, this emission most likely arises from irregularly distributed galactic cirrus in this direction; this is supported by the relatively lower quality IRAS 100 $\mu$m image, which shows substantial extended cirrus beyound our survey region. The ISOPHOT observation of a possible shell at 300 arcseconds was based on a $53^\prime \times 3^\prime$ linear scan with $30^{\prime\prime} \times 92^{\prime\prime}$ pixels, at a position angle of 8 degrees east of north centered on the source, which in our case would be sampling the emission in the northern scan. We see with better azimuthal coverage with MIPS that the emission in both of our scans is likely strongly contaminated with galactic cirrus. The presence of the inner shell reported by Speck et al. at 150 arcseconds cannot be directly confirmed by our observations because the 160 $\mu$m PSF Airy ring at 170 arcseconds from the source overlaps that region. The surface brightness at the Airy ring is about 17 MJy Sr$^{-1}$ after background subtraction (but before the PSF subtraction). This is comparable to the background subtracted surface brightness of 20 (30) MJy Sr$^{-1}$MJy Sr$^{-1}$ at 120 (180) $\mu$m reported by \citet{2000ApJ...545L.145S}. 

\begin{figure}
	\centering
        \plotone{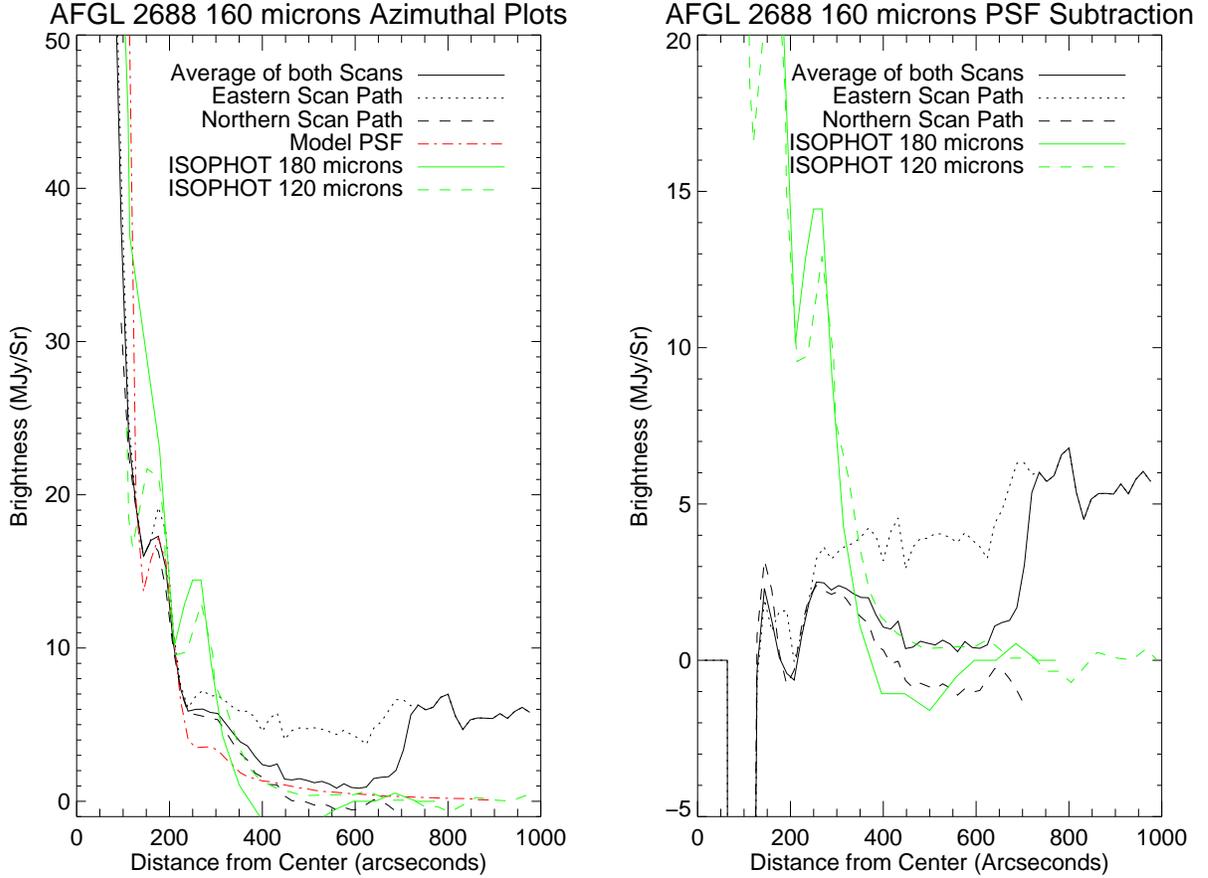}
	\caption{\textbf{Left}: Azimuthally averaged surface brightness of AFGL 2688 at 160 $\mu$m. The two scan paths of the observations are plotted separately to show the differences in their background levels. The solid line represents the average of the two scans. Beyond about 750 arcseconds from the source, the northern scan path stops so only the eastern path is represented beyond this distance. The PSF is scaled to the brightness of the Airy ring at 170 arcseconds from the source. \textbf{Right}: The result of PSF subtraction. The region less than 200 arcseconds from the source may be affected by the PSF subtraction so we cannot make  strong quantitative statements about the emission there at this time, except that the surface brightness of this region is likely to be about 2 MJy Sr$^{-1}$. ISOPHOT data from \citet{2000ApJ...545L.145S} are also plotted for comparison after subtracting the background offset determined from the median of the values beyond about 400 arcseconds from the source. The reported ISOPHOT intensity at 300 arcseconds is almost eight times higher than the surface brightness seen by MIPS at 160 $\mu$m after PSF subtraction.}
	\label{fig:afgl2688_160_1d}
\end{figure}

In the azimuthally averaged surface brightness of our 160 $\mu$m image (Figure \ref{fig:afgl2688_160_1d}), we see a similar result as the 70 $\mu$m data. Speck et al. reported a brightness value of the putative shell at 300 arcseconds from the source to be 20 MJy Sr$^{-1}$ at 120 $\mu$m and 60 MJy Sr$^{-1}$ at 180 $\mu$m, without background subtraction. The 120 $\mu$m emission is about 10 MJy Sr$^{-1}$ above their background as extrapolated from the surface brightness far from the star in their plot. At 180 $\mu$m, the excess emission is about 15 MJy Sr$^{-1}$ above the background. In the MIPS 160 $\mu$m image, we find that the surface brightness at 300 arcseconds from the source is about 2 MJy Sr$^{-1}$ after background subtraction. Although the comparison is not at the same wavelength, the excess emission we detect is 7 times below that measured by ISO. When comparing the two scans in our observations separately, we see that the eastern scan has roughly uniform brightness of 4 MJy Sr$^{-1}$ out to 1000 arcseconds, while the emission in the northern scan drops off at 400 arcseconds from the source from about 2 MJy Sr$^{-1}$ to the level we use as the background in the data reduction. Both the asymmetry in the enhanced emission in both scans and the low surface brightness in the Spitzer data at 300 arcseconds from the source lead us to conclude that the shell at this distance reported on the basis of ISO data is probably not associated with the source.

\subsection{OH 231.8+4.2}

\begin{figure}
	\centering
	\plotone{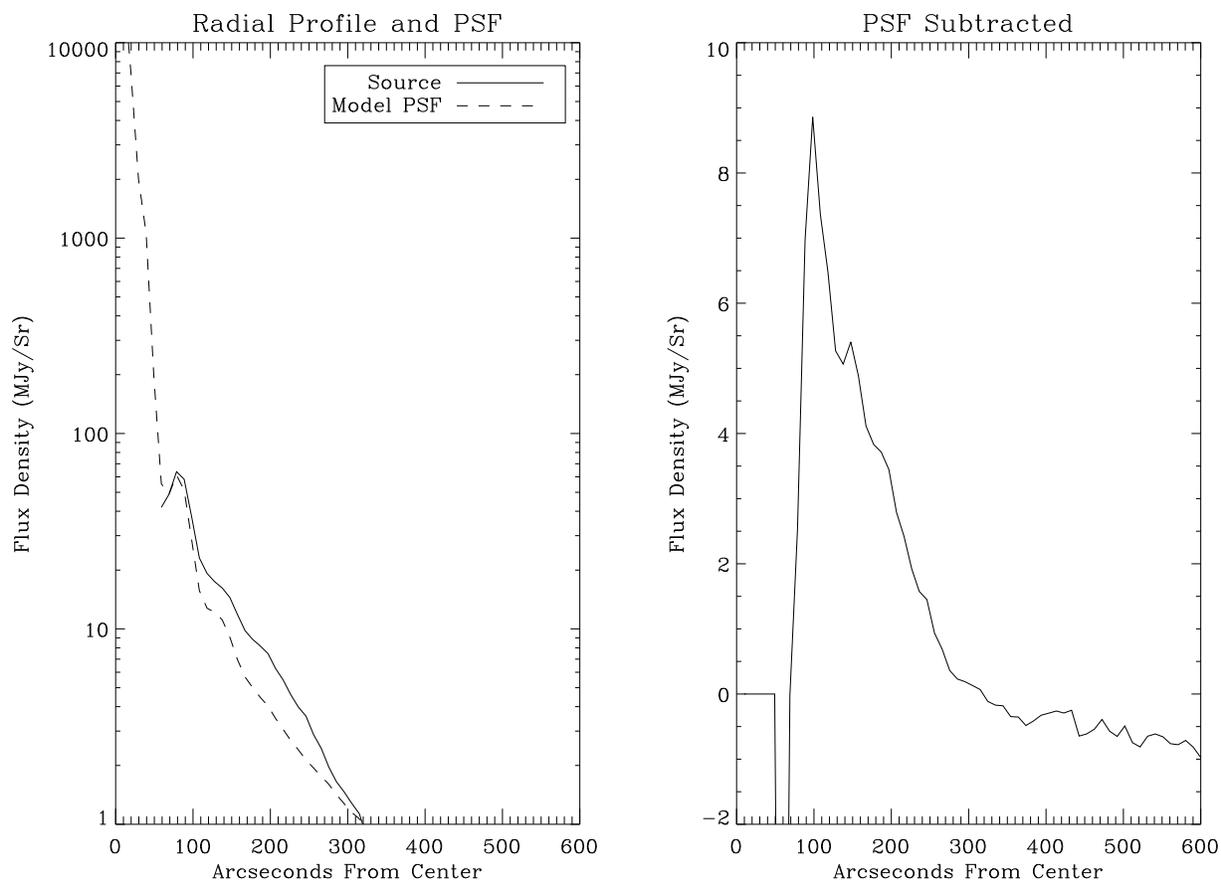}
	\caption{Left: the azimuthally averaged surface brightness of OH 231.8 and the model PSF at 70 $\mu$m. The model PSF has been scaled to match the Airy ring emission of OH 231.8 at a radius of 76 arcseconds. Right: the result of subtraction of the model PSF. The shape of the residuals is almost the same as those left from the PSF subtraction of AFGL 2688 (Figure \ref{fig:afgl2688_70_1d}), falling off to background levels beyond 250 arcseconds from the source.}
	\label{fig:oh231_70_1d}
\end{figure}

Neither the 70 nor 160 $\mu$m images of OH 231.8 show any signs of spherically symmetric extended emission. The PSF subtraction at 70 $\mu$m leaves similar residuals as for AFGL 2688, though the brightnesses of these residuals are much lower because the central source is intrinsically less bright. The strongest PSF residuals are along the diffraction spikes, similar in shape to those from AFGL 2688. This further suggests that these residuals are not physical features, but are rather PSF artifacts not accounted for by the \textit{stinytim} model.  In the 160 $\mu$m azimuthally averaged plots, both scans appear to have rather uniform emission, with the northern scan having slightly higher surface brightness at approximately 2 MJy Sr$^{-1}$, while the eastern scan has an average surface brightness of 1 MJy Sr$^{-1}$ (Figure \ref{fig:oh231_70_1d}). 

\begin{figure}
	\centering
	\plotone{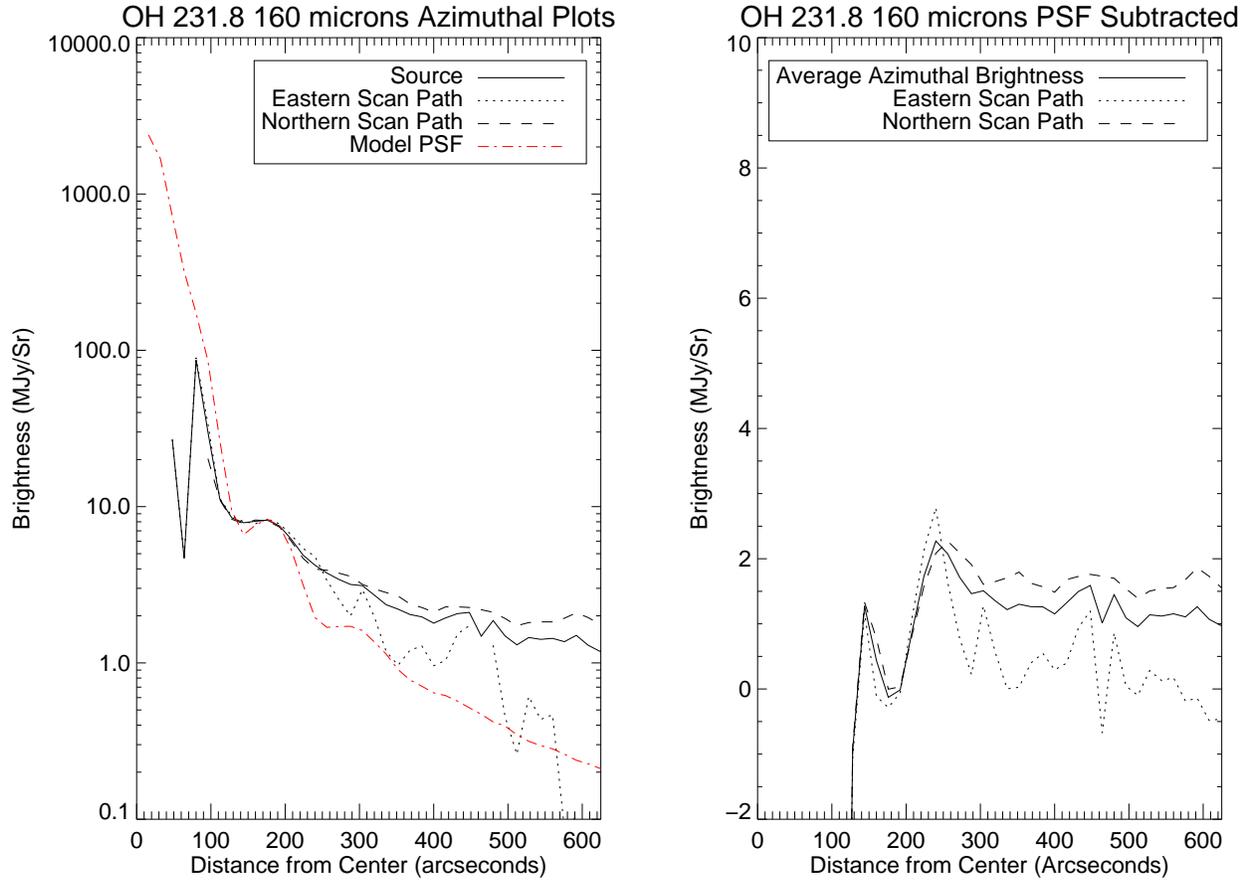}
	\caption{Left: the azimuthally averaged surface brightness of OH 231.8 along with the model PSF at 160 $\mu$m. Right: the 1-D PSF subtraction. There is no apparent enhanced emission other than attributable to fluctuations from the galactic cirrus.}
	\label{fig:oh231_160_1d}
\end{figure}

The 160 $\mu$m image shows some slightly clumpy emission along the northern scan path, slightly beyond the PSF diffraction spike (see Figure \ref{fig:initial_mosaic}) at 320 arcseconds. The clumps, with a width of about 80 arcseconds, have a surface brightness of about 3.6 MJy Sr$^{-1}$ on a 2 MJy Sr$^{-1}$ diffuse background. The IRAS 100 $\mu$m image shows diffuse galactic cirrus in this region, which may be the origin of these clumps. They are aligned with the bipolar axis, but unfortunately also with the diffraction ray so it may also be a PSF feature. Since this study is concerned with investigating spherically symmetric emission, we will defer the question of their nature.  The azimuthally averaged surface brightness shows no enhanced emission attributable to the source (Figure \ref{fig:oh231_160_1d}).

\subsection{IRAS 16342$-$3814}

In contrast to the previous two sources, IRAS 16342$-$3814 does show some evidence for what may be a large patchy shell at both 70 and 160 $\mu$m.  The enhanced emission out to a radial distance of $\sim$ 400 arcseconds is consistent with rough circular symmetry about the star in the coverage area available (Figure \ref{fig:initial_mosaic}). This radius corresponds to a physical size of about 4 parsecs, assuming the distance to IRAS 16342$-$3814 to be 2 kpc \citep{1999ApJ...514L.115S}. This possible shell structure is rather patchy, with a large arc in the western scan at 70 $\mu$m. At 160 $\mu$m, the same arc is present, but is more diffuse. Nevertheless, the azimuthal average shows an excess above the background at $\sim$ 400 arcseconds from the central source. The brightest portion of this feature occurs at a position angle of 250 degrees east of north, which is the same direction as the axis of the bipolar nebula seen in the optical. The ratio of the 70 to 160 $\mu$m average surface brightness, is roughly constant ($\sim 0.3$) as a function of radial distance from the source. This ratio yields a color temperature of $\sim 32 \pm 2$ K, and, if a $\nu^1$ emissivity law is assumed, a dust temperature of $26 \pm 2 $ K. The constant color temperature suggests that the dust is heated primarily by the interstellar radiation field.

\begin{figure}
	\centering
	\plotone{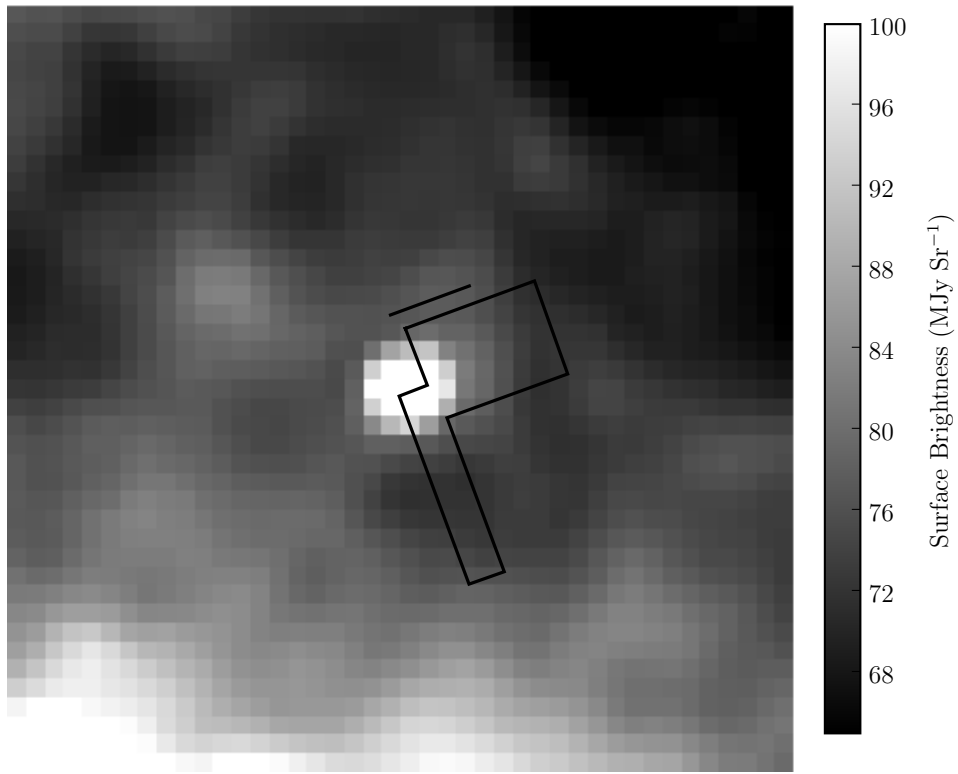}
	\caption{The MIPS 70 $\mu$m field of view of IRAS 16342$-$3814 overlaid on the IRAS 100 $\mu$m image of the surrounding regions. Note the abundance of galactic cirrus. The bar above the frame is 400 arcseconds in length, representing the radius of the potential shell seen with MIPS around IRAS 16342$-$3814.}
	\label{fig:iras16342_iras100}
\end{figure}

Because there are a number of interstellar cirrus features in the general direction of IRAS16342-3814, we cannot definitively argue that the patchy arc of emission in the scan paths around this source is caused by mass loss from the star. The IRAS 100 $\mu$m image of this region with an overlay of the coverage of our data (figure \ref{fig:iras16342_iras100}) shows that there is indeed a substantial likelihood that this emission is cirrus. With this caveat in mind, we proceed in the analysis below on the assumption that the extended emission has resulted from mass loss from the star to obtain some indicative numbers for the mass loss rate necessary to produce such a shell. In the following analysis, we will use the term ``mass outflow rate'' to indicate the total mass flowing outward per unit time across a spherical shell at a particular radius; this includes both the stellar component and the interstellar medium being swept up; we will use the term mass loss rate to refer to mass loss from only the star.

\begin{figure}
	\centering
	\plotone{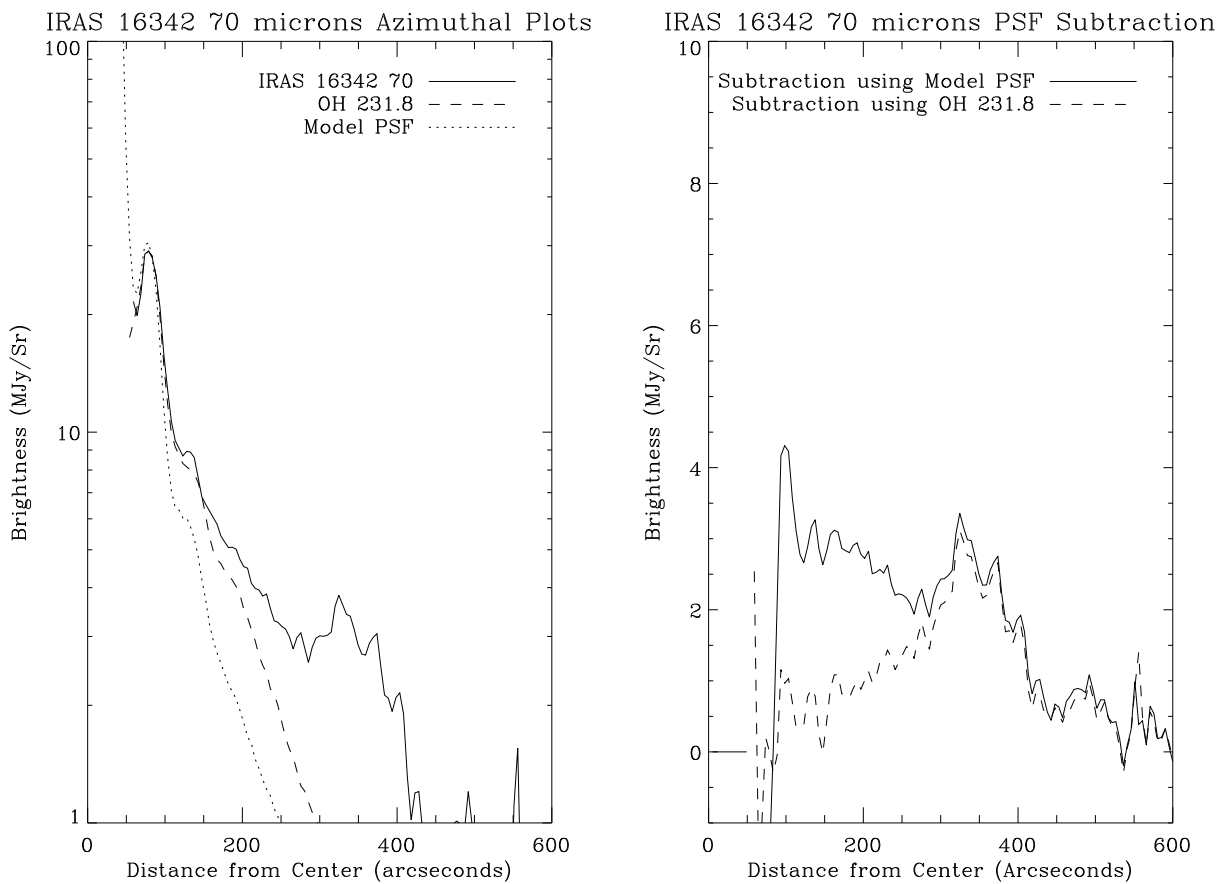}	
	\caption{Azimuthal average and PSF subtraction of IRAS 16342$-$3814 at 70 $\mu$m. \textbf{Left}: As with Figure \ref{fig:afgl2688_70_1d}, this plot shows the source plotted along with the radial profiles of OH 231.8 and the model PSF. Unlike AFGL 2688, IRAS 16342$-$3814 shows a definite excess in its radial profile above the background and PSF profile. \textbf{Right}: the difference between using the model and OH 231.8 for the PSF subtraction. Both give about the same level of excess emission of $\sim$ 2.5 MJy Sr$^{-1}$ between 300 to 400 arcseconds from the source, where we see a patchy shell in the MIPS 70 and 160 $\mu$m images.}
	\label{fig:iras16342_70_1d}
\end{figure}

\begin{figure}
	\centering
	\plotone{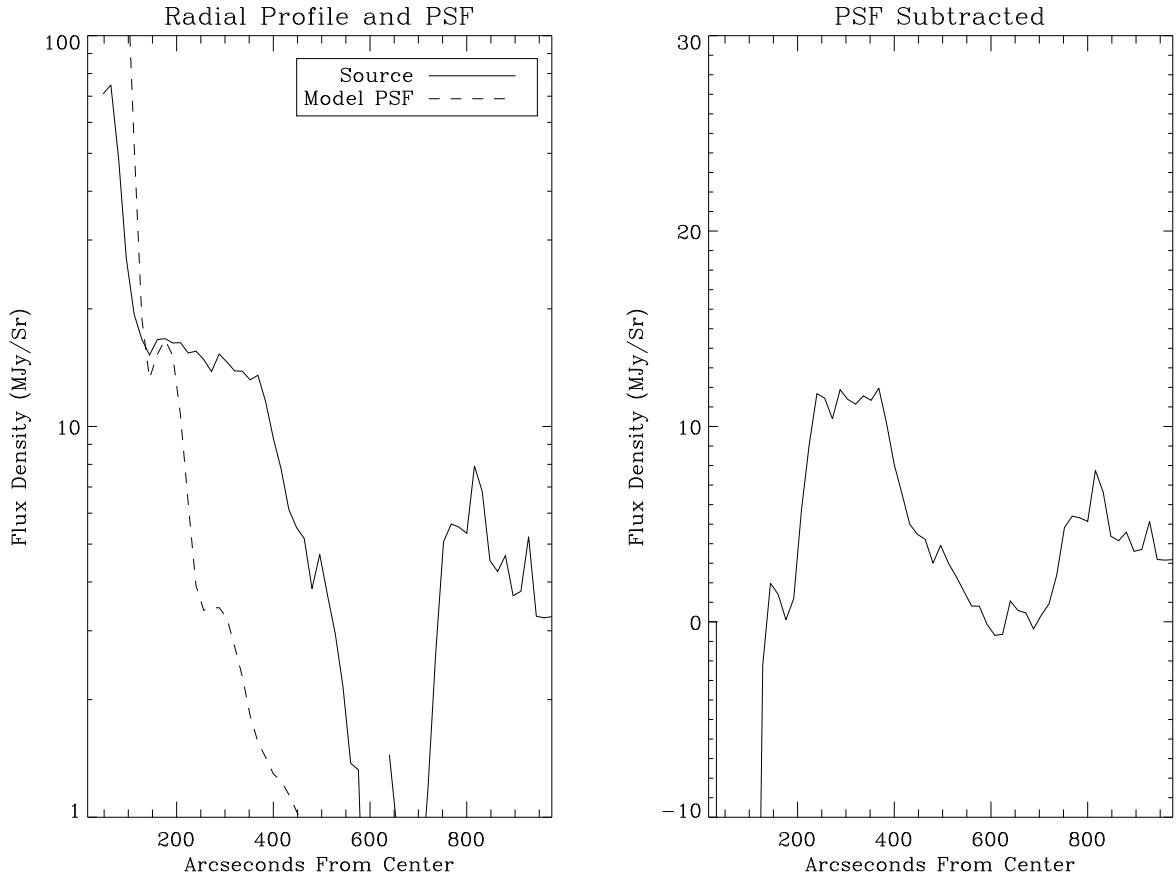}
	\caption{Left: The azimuthally averaged surface brightness of IRAS 16342$-$3814 and the model PSF scaled to the brightness of the source at the location of the 170 arcsecond radius Airy ring of the 160 $\mu$m PSF. Right: the result of PSF subtraction. The excess emission seen in the MIPS images is clearly present out to 400 arcseconds.}
	\label{fig:iras16342_160_1d}
\end{figure}

\begin{figure}
	\centering
	\includegraphics[scale=.7,angle=90]{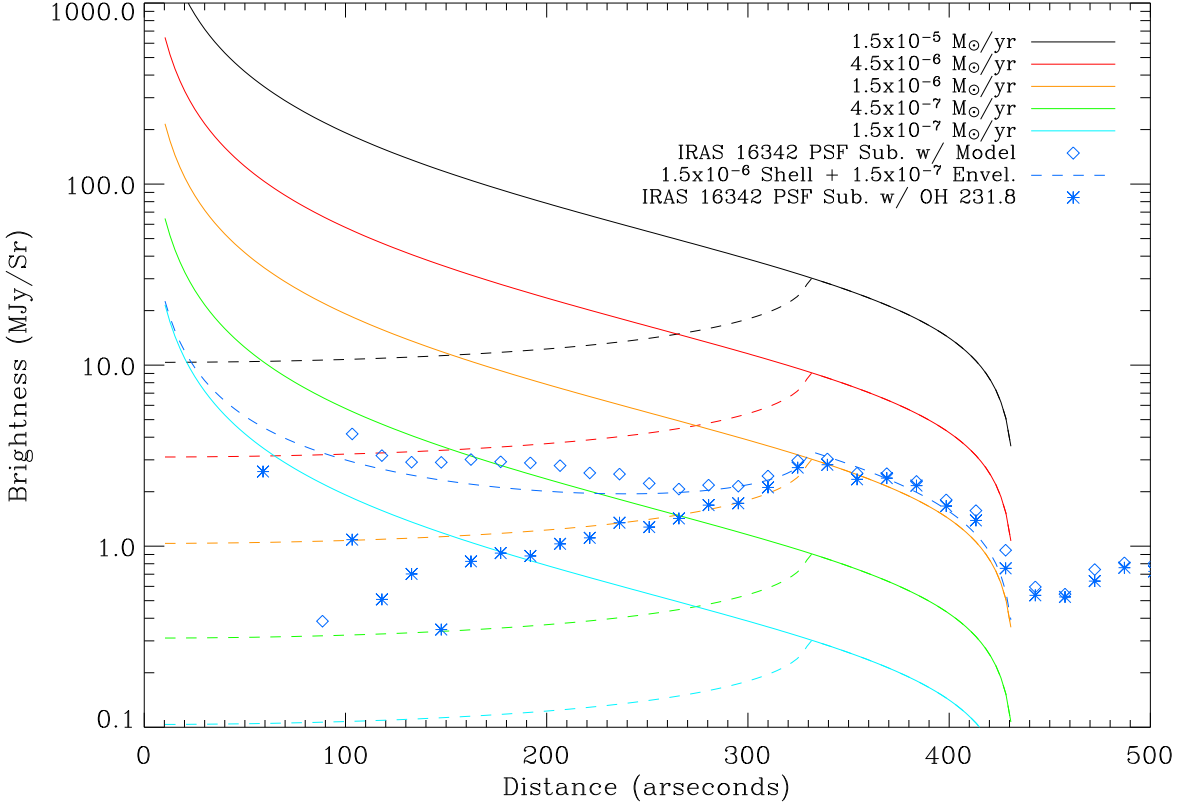}
	\caption{Plot of the expected surface brightness for various mass loss rates. The solid line shows the expected surface brightness for a smooth constant mass loss envelope with a radius of 4.2 pc while the dashed lines show the radial profile of a single shell of 1 pc thickness at the same mass loss rate. The diamonds show the radial profile of IRAS 16342$-$3814 after the subtraction of the model PSF. We find that the best fit to the radial brightness profile obtained from subtracting the model PSF is a two-component model with a shell of radius 4.2 pc and thickness 1 pc requiring a dust mass loss rate of $10^{-6}$ $M_{\odot}$~ yr$^{-1}$ superimposed upon an envelope from a constant dust mass loss of $10^{-7}$ $M_{\odot}$~ yr$^{-1}$ (dashed blue line). The asterisks show the radial profile found from using OH 231.8 as a substitute for the model PSF in the PSF subtraction will fit using only the shell component with a dust mass loss rate of $10^{-6}$ $M_{\odot}$~ yr$^{-1}$ (dashed orange line). Points closer than a radial distance of $\sim$ 150 arcseconds are unreliable because they are most affected by PSF subtraction residuals and are not used in the fit.}
	\label{fig:iras16342_model}
\end{figure}

Making a few assumptions about the mass outflow and the dust, we can set some limits on the mass outflow rate of IRAS 16342$-$3814. Assuming that the mass loss is spherically symmetric, we can fit a model of the expected brightness profile to the observed radial profile of the source. Figures \ref{fig:iras16342_70_1d} and \ref{fig:iras16342_160_1d} show the azimuthally averaged radial profiles at 70 and 160 $\mu$m, respectively. We fit only the 70 $\mu$m radial profile because there is less contamination by galactic cirrus compared to 160 $\mu$m. Following the derivation by \citet{1986ApJ...310..842G}, if we assume a $1/r^2$ density profile, appropriate for constant mass outflow, where $r$ is the distance from the star, and a constant expansion velocity, $v_e$, then the surface brightness $I_\nu(b)$ has the form: 

\begin{eqnarray}
I_\nu(b) = \frac{\stackrel{.}{M}\kappa_\nu B_\nu(T)}{2\pi v_e b}\cos^{-1}(\frac{b}{R_{max}}) \quad\mbox{for} ~ R_{min} \leq b \leq R_{max} \nonumber \\
I_\nu(b) = \frac{\stackrel{.}{M}\kappa_\nu B_\nu(T)}{2\pi v_e b}\left[\cos^{-1}(\frac{b}{R_{max}}) - \cos^{-1}(\frac{b}{R_{min}})\right]  \quad\mbox{for} ~ b < R_{min}
\label{eq:shell_model}
\end{eqnarray}
where $b$ is the projected physical distance from the source in the sky, $\kappa_\nu$ is the opacity coefficient in units of cm$^2$ g$^{-1}$, $B_\nu(T)$ is the blackbody function, and $R_{max}$ and $R_{min}$ are the outer and inner radii of the shell. We further assume that the outflow velocity is constant at 15 km s$^{-1}$, and that the dust is composed of small astronomical silicate grains with $\kappa_\nu$ = 104 cm$^2$ g$^{-1}$ at 70 $\mu$m (using a = 0.1 $\mu$m, $\rho = 2.3$ g cm$^{-3}$, and $Q_{abs}^\nu = 2.99\times10^{-3}$ from \citet{1984ApJ...285...89D}). We also adopt a constant temperature of 26 K throughout the shell and envelope. Figure \ref{fig:iras16342_model} gives a comparison between different models having various mass outflow rates. The best-fit model is a shell with a 4.2 pc radius and a thickness of 1 pc, with a dust outflow rate of $1.5\times10^{-6} M_{\odot}$ yr$^{-1}$ superimposed on a smooth envelope with a constant dust outflow rate of $1.5\times10^{-7} M_{\odot}$ yr$^{-1}$. This implies a rather high gas outflow rate on the order of $3\times10^{-4} M_{\odot}$ yr$^{-1}$ in the shell and $3\times10^{-5} M_{\odot}$ yr$^{-1}$ in the smooth envelope, assuming a gas to dust mass ratio of 200. Using OH 231.8 as the empirical PSF, the fit only requires a shell component with a mass outflow rate of $3\times10^{-4} M_{\odot}$ yr$^{-1}$ without an underlying smooth envelope. These fits depend on the choice of PSF, but once the PSF is chosen, the mass outflow rates are insensitive to residuals from PSF subtraction, since we only fit for the region beyond 150 arcseconds from the central source (see Figure \ref{fig:iras16342_70_1d}). However, our value for the mass outflow rate is highly dependent on the assumed temperature of the dust. For typical ISM dust temperatures between 22 and 35 K, the inferred dust outflow rate is related empirically to the adopted temperature by: $\stackrel{.}{M} ~ \propto ~ T^{-7.6}$ at 70 $\mu$m. If T $ >$ 26 K, we would infer a substantially lower mass outflow rate.

Given the model thickness of the shell and assuming a typical expansion velocity of 15 km s$^{-1}$, we estimate the maximum duration of the enhanced mass loss event which produced the shell to be about 65,000 years. This is an upper limit because the internal velocity dispersion in the shell broadens the shell as it expands. A 1 km s$^{-1}$ dispersion, for example, would reduce the duration of the mass loss that produced the shell by about a third. If we use an age of 40,000 years for the duration of  the enhanced mass loss and approximate the shell as spherical, the total dust mass would be about 0.04 $M_{\odot}$, giving a total shell mass of $\sim$ 8 $M_{\odot}$, assuming a gas-to-dust mass ratio equal to 200. The velocity of the shell is probably slower than the molecular outflow velocity of a typical AGB star and the total amount of mass lost by the AGB star is likely smaller than this amount because of the interaction with the ISM out to the distances we are observing (see below). 

The mass of the shell can be estimated directly from the infrared emission using the equation $ M = F_\nu D^2/(B_\nu(T)\kappa_\nu)$, where $F_\nu$ is the total flux density in the shell and $D$ is the distance to the source. The integrated flux density from the limited azimuthal coverage that we have between radii of 300 and 400 arcseconds is 6.4 Jy at 70 $\mu$m, corresponding to a dust mass of 0.015 $M_\odot$ in our observed portion of the shell, assuming a temperature of 26 K for the dust. We have about one fifth of the full azimuthal coverage with data at this distance from the star, and can estimate the total dust mass of the shell if we approximate the shell as isotropic with an averaged surface brightness of 2.5 MJy Sr$^{-1}$ at 70 $\mu$m from the radial profile, with a width of 100 arcseconds, and a temperature of 26 K. The total dust mass of the extrapolated shell is then about 0.03 $M_{\odot}$, comparable to the above estimate. 

Because the shell is so large, the amount of interstellar matter that may have been swept up and accumulated in the shell during the AGB could be significant. For an ISM hydrogen density of 1 cm$^{-3}$, a 4 parsec radius shell would have swept up about 7 $M_{\odot}$, which would account for almost all the mass that we may be measuring. The accumulation of interstellar material could potentially lower the stellar mass loss rate estimated from our model by a large factor. We therefore emphasize that the mass outflow rate estimate is only an upper limit to the stellar mass loss rate.

One of the difficulties in addressing the mass loss history of IRAS 16342$-$3814 with the current data is its location in the Galaxy; with a scale height of only 150 pc, there is possible confusion with diffuse galactic cirrus. An unfortunate alignment with background emission is an alternative to the existence of a shell produced by mass loss. Further observations with better azimuthal coverage are needed to test the circumstellar shell hypothesis.

\section{Discussion}

The extended dust emission in AFGL 2688 and OH 231.8 seems to show that these two objects do not have as long a mass loss history as one might have anticipated. Our results for these two sources probe the region beyond $\sim$ 100 arcseconds from the central star, which corresponds to $1.0 \times 10^4$ yrs ago for AFGL 2688 and $2.8 \times 10^4$ yrs ago for OH 231.8, given a distance of 420 pc \citep{2006ApJ...641.1113U} and 1.2 kpc \citep{1985ApJ...292..487J}, respectively, and an expansion velocity of 20 km s$^{-1}$. We can can set a limit to the dust mass loss rate by inverting equation \ref{eq:shell_model}:
\begin{equation}
\dot{M}  = \frac{I_\nu(b, T) 2\pi v_e b }{\kappa_\nu B_\nu(T) \cos^{-1}(b/R_{max})}
\label{eq:mdot_limit}
\end{equation}
The dust mass loss rate can be simplified as a function of the observed surface brightness $I$ and the projected distance from the source in the sky $ b $ with the assumption that $ b \ll R_{max}$: 
\begin{eqnarray*}
\dot{M}(70 \mu m) & \sim & 3.8\times10^{-8} M_\odot ~ yr^{-1} \left(\frac{I_{70 ~\mu m}}{1 ~ \hbox{MJy Sr$^{-1}$}}\right) \left(\frac{100 ~ cm^2 g^{-1}}{\kappa_{70 ~ \mu m}}\right)\left(\frac{b}{1 ~ pc}\right) \left(\frac{v_e}{15 ~ km ~s^{-1}} \right) \left(\frac{T}{30 ~ K}\right)^{-7.6} \\
\dot{M}(160 \mu m) & \sim & 2.9\times10^{-8} M_\odot ~ yr^{-1} \left(\frac{I_{160 ~\mu m}}{1 ~ \hbox{MJy Sr$^{-1}$}}\right) \left(\frac{20 ~ cm^2 g^{-1}}{\kappa_{160 ~ \mu m}}\right) \left(\frac{b}{1 ~ pc}\right) \left(\frac{v_e}{15 ~ km ~s^{-1}} \right) \left(\frac{T}{30 ~ K}\right)^{-3.6}
\end{eqnarray*}

If $b \sim R_{max}$, then the full form of equation \ref{eq:mdot_limit} must be used because the $\cos^{-1}(b/R_{max})$ term in the denominator of equation \ref{eq:mdot_limit} becomes important and will cause the dust mass loss rate inferred from a given surface brightness value to increase drastically. For the cases such as AFGL 2688 and OH 231.8 where we do not see a clear envelope associated with the source, we cannot be sure that the condition $b \ll R_{max}$ holds since $R_{max}$ is indeterminate. The equations above also show the temperature dependence in the form of a power law to approximate the blackbody function between 22 and 35 K. From these equations we see that the temperature dependence at 160 $\mu$m is less steep than at 70 $\mu$m, but the contamination by galactic cirrus near the sources makes estimating upper limits problematic at 160 $\mu$m. We can establish an upper limit to the mass loss rates for AFGL 2688 and OH 231.8 at 70 $\mu$m based up on the surface brightness of the residual left from the PSF subtraction at 200 arcseconds from the source, which is a compromise between a location far enough from the source that PSF subtraction errors are small and being close enough to the source that there could plausibly be a circumstellar envelope. For AFGL 2688, 200 arcseconds corresponds to a radial distance from the source of 0.4 parsecs, using a distance to the source of 420 parsecs \citep{2006ApJ...641.1113U}. The surface brightness after PSF subtraction with the model PSF is about 14 MJy Sr$^{-1}$ at 70 $\mu$m. Assuming an expansion velocity of 20 km s$^{-1}$ as observed in CO by \citet{1997A&A...328..290S}, a dust temperature of 30 K, $\kappa_\nu$ = 104 cm$^2$ g$^{-1}$ appropriate for carbon dust \citep{1984ApJ...285...89D}, and that the possible envelope has a radius significantly greater than 0.4 pc, then the dust mass loss rate upper limit is $2.1\times10^{-7} $ M$_\odot$ yr$^{-1}$. Similarly for OH 231.8, the residual emission from the PSF subtraction is 3 MJy Sr$^{-1}$ at 200 arcseconds from the central star, which corresponds to 1.3 parsecs at a distance of 1.3 kpc. This surface brightness implies a dust mass loss rate of $1.0\times10^{-7}$ M$_\odot$ yr$^{-1}$ assuming an outflow velocity of 15 km s$^{-1}$, $\kappa_\nu$ = 98 cm$^2$ g$^{-1}$ (appropriate for silicate dust), and a temperature of 30 K.

There are several potential explanations for the lack of extended emission seen in this study. One is that, even with Spitzer's increased sensitivity, the dust emission is below the threshold for detection. Since the distances we are studying are far from the central star, we would expect the temperature of the dust to be determined by the ambient interstellar radiation field. If the dust temperature is about 20 K, typical for the ISM \citep{1983A&A...128..212M}, and an emissivity $\propto \nu^1$, the surface brightness would be 12 times stronger at 160 $\mu$m, than at 70 $\mu$m where the resolution and sensitivity of MIPS is better. We would be more likely to detect cool dust emission at 160 $\mu$m than at 70 $\mu$m. However, confusion with the galactic cirrus is also likely at 160 $\mu$m because it is about the same temperature as any hypothetical extended dust emission associated with the source that is heated primarily by the interstellar radiation field. For comparison, our detection of the possible shell from IRAS 16342-3184 has a ratio of $I_{160 \mu m}/ I_{70\mu m} \sim 3.3$, corresponding to a color temperature of 32 K, which requires a higher than average interstellar radiation field. 

The upper limits to the mass loss rates derived in this study can be compared with theoretical AGB evolutionary models, particularly those that predict enhanced mass loss rates due to thermal pulsation near the end of the AGB. The spatial coverage in this study between $\sim$ 200 to 1000 arcseconds from the central star translates into a probe of the history of mass loss between $2\times10^4$ yrs to $1\times10^5$ years ago for AFGL 2688 and $6\times10^4$ yrs to $3\times10^5$ yrs ago for OH 231.8, using an expansion velocity of 20 km s$^{-1}$ for both sources. The models by \citet{1993ApJ...413..641V} show that for a 2.0 M$_\odot$ progenitor, during the last few $\times10^5$ yrs of AGB evolution, there are several thermal pulses which result in enhanced mass loss rates peaking at about $1.3\times10^{-5}$ M$_\odot$ yr$^{-1}$ during the end of each pulse. Using a gas to dust mass ratio of 200, we find that for AFGL 2688 and OH 231.8, the upper limit to the total mass loss rate during the above time intervals is about $4\times10^{-5}$ M$_\odot$ yr$^{-1}$ and $2\times10^{-5}$ M$_\odot$ yr$^{-1}$, respectively. These limits are close to the sensitivity necessary to see shells that may be the result of thermal pulsation on the AGB. Although we do not detect the shells reported by \citet{2000ApJ...545L.145S}, which they attribute to thermal pulses, the signatures of thermal pulses may still be present, but below our current sensitivity.

We also consider the possibility that OH 231.8 does not show extended emission in the far infrared because of interactions with a binary companion, which caused the central star to lose mass more rapidly and more recently than during the evolution of a lone AGB star. A companion to QX Pup (the central star of OH 231.8), is evidenced by optical spectra consistent with a companion of stellar type A0 V \citep{2004ApJ...616..519S}. QX Pup also shows a peculiar paradox of being an M9III \citep{1981PASP...93..288C} AGB star, while its bipolar activity and morphology display all the signs of post-AGB activity of typical PPNe. Having a close companion would enhance the mass loss rate and provide a mechanism for generating the bipolar outflows \citep{1987PASP...99.1115M}, thus shortening its mass loss history enough that we should not be surprised to see no emission far from the source. In contrast to the collimated outflow from OH 231.8 \citep{2001A&A...373..932A}, AFGL 2688 has a spherically symmetric envelope seen in $^{13}$CO \citep{1996ApJ...465..926Y} and evidenced by the partial, concentric, circular arcs present in HST images \citep{1998ApJ...492L.163S}. Since a binary interaction would not cause the past spherically symmetric mass loss, a possible companion is probably not a good explanation for initiating mass loss, although the present bipolar morphology of AFGL 2688 is consistent the possibility that a binary interaction has altered the mass loss in more recent times.

Tracing the mass loss history during the AGB phase should be relatively straightforward via mapping the emission from the circumtellar envelope. But in practice, it has been difficult because molecular-line observations are ultimately limited by the photodissociation of molecules in the outer envelope regions, and far-infrared observations of dust emission have been limited by the generally low angular resolution of the space-based telescopes which have been available for this purpose in the past (IRAS and ISO). Hence the reported detections of very extended emission in a few dying stars with IRAS and ISO has generally been recognized as an important milestone in the study of mass loss. However, such detections have also raised the very important question of why only a few select objects like Y CVn \citep{1996A&A...315L.221I} or RY Dra \citep{1993ApJ...409..725Y} -- not particularly known for having high mass loss rates -- show extended envelopes, whereas large numbers of stars having high CO-determined mass loss rates do not reveal the presence of such envelopes. Is this because the radial density law for most of these high mass-loss stars is significantly steeper than in objects like Y CVn or RY Dra, and in particular steeper than $r^{-2}$ (an issue of profound importance for the evolutionary times of stars through the AGB phase and theories of mass-loss), or are the claimed detections of extended envelopes really a result of poorly characterised instrumental artifacts? The Spitzer data presented in this paper clearly show that the presence of these shells cannot be confirmed at a level well below the intensities expected from Speck et al.'s results. Our non-detections call into question not only the ISO results on AFGL 2688, but all results on the detection of extended envelopes in other objects using the same linear scan technique described by Speck et al.. More detailed mapping of many high mass-loss rate objects like AFGL 2688, to search for such shells, is crucially needed.

\section{Summary}

Spitzer observations of extended envelopes of expanding, dusty outflows from bipolar AGB and post-AGB stars reveal that there may be a very large dust shell having a radius of 400 arcseconds around IRAS 16342$-$3814. The combination of the presence of nearby cirrus emission and the limited azimuthal coverage in the images makes the conclusion for a shell uncertain, but if the shell is indeed the result of an episodic mass loss event, then it would represent one of the largest dust shells found so far. 

Our observations of AFGL 2688 at 70 $\mu$m do not show the dust shells at 150 and 300 arcseconds from the source reported by \citet{2000ApJ...545L.145S}. Since the dust shell may be very cool, the non-detection at 70 $\mu$m does not alone rule out dust shells. However, we find that there is only a slight excess at 160 $\mu$m of 2 MJy Sr$^{-1}$ above the background at the reported location of the outer shell ($\sim 300$ arcseconds). With greater azimuthal coverage than was previously obtained with ISOPHOT we find that there is substantial contamination by galactic cirrus in the region at 160 $\mu$m, with galactic cirrus emission above the sky background present throughout the entire eastern scan path of our observations. Unfortunately, at 160 $\mu$m we can only set an upper limit to the emission from a dust shell at 150 arcseconds from the source because one of the Airy rings of the PSF overlaps that region. We also see no extended emission from OH 231.8 at either 70 or 160 $\mu$m other than that attributable to galactic cirrus. In fact, using OH 231.8 as an empirical point source for PSF subtraction from AFGL 2688 at 70 $\mu$m appears to substantially reduce PSF subtraction residuals compared to using the model PSF.

The limitation of our method of observation is that we have only two radial directions to probe possible extended emission. For cases like IRAS 16342$-$3814 where the shell is patchy, better azimuthal information would help to resolve whether the origin of the emission is from the star or from galactic cirrus. The MIPS data show that observing very extended emission in the far infrared is possible, but difficult because of the prominence of galactic cirrus emission at these wavelengths.

This work is based on observations made with the Spitzer Space Telescope, which is operated by the Jet Propulsion Laboratory, California Institute of Technology under a contract with NASA. Support for this work was provided by NASA through an award issued by JPL/Caltech. MM and RS was partially funded for this work from a GO Spitzer award and an LTSA award (no. 399-20-40-06) from NASA.

\clearpage

\begin{table}
\centering
\begin{tabular}{ccccc}
\hline\hline
 & Mean Before &  & Mean After &  \\ 
 & Subtraction & $\sigma$ & Subtraction & $\sigma$ \\ 
Object & (MJy Sr$^{-1}$) & (MJy Sr$^{-1}$) & (MJy Sr$^{-1}$) & (MJy Sr$^{-1}$) \\ \hline
AFGL 2688 & 14.1 & 1.1 & 0.6 & 0.9 \\ 
OH 231.8 & 10.2 & 0.9 & 0.1 & 0.6 \\ 
IRAS 16342$-$3814 & 30.3 & 1.3 & 0.3 & 0.9 \\ \hline
\end{tabular}
\caption{The 70 $\mu$m mean surface brightness and sensitivity for a region of uniform background far from the source, before and after the median background removal described in the text.}
\label{tab:70_sensitivity}
\end{table}

\clearpage

\begin{table}
\centering
\begin{tabular}{ccccc}
\hline\hline
 & Mean Before &  & Mean After &  \\ 
 & Subtraction & $\sigma$ & Subtraction & $\sigma$ \\ 
Object & (MJy Sr$^{-1}$) & (MJy Sr$^{-1}$) & (MJy Sr$^{-1}$) & (MJy Sr$^{-1}$) \\ \hline
AFGL 2688 & 31.4 & 1.7 & 0.3 & 1.5 \\ 
OH 231.8 & 23.1 & 0.7 & -0.2 & 0.7 \\ 
IRAS 16342$-$3814 & 90.1 & 2.9 & 1.0 & 2.3 \\ \hline
\end{tabular}
\caption{The 160 $\mu$m mean surface brightness and sensitivity for a region of uniform background far from the source, before and after the median background removal described in the text.}
\label{tab:160_sensitivity}
\end{table}

\clearpage

\end{document}